# MesoNet: A Fundamental Principle for Multi-Representation Learning in Complex Chemical Systems

Jinming Fan, Chao Qian and Shaodong Zhou*

[a] *College of Chemical and Biological Engineering, Zhejiang Provincial Key Laboratory of Advanced Chemical Engineering Manufacture Technology, Zhejiang University, 310027 Hangzhou (P. R. China)*

[b] *Institute of Zhejiang University – Quzhou, 324000 Quzhou (P.R. China)*

*E-mail:* szhou@zju.edu.cn

*Supporting information and the ORCID identification numbers for the authors of this article can be found under: (Zhou: 0000-0003-3048-4678)*

***Abstract:*** *Accurate prediction of molecular properties in complex chemical systems is crucial for accelerating material discovery and chemical innovation. However, current computational methods often struggle to capture the intricate compositional interplay across complex chemical systems, from intramolecular bonds to intermolecular forces. In this work, we introduce MesoNet, a novel framework founded on the principle of multi-representation learning and specifically designed for multi-molecule modeling. The core innovation of MesoNet lies in the construction of context-aware representation – dynamically enriched atomic descriptors generated via Neural Circuit Policies. These parameters efficiently capture both intrinsic atomic properties and their dynamic compositional context through a cross-attention mechanism spanning both intramolecular and intermolecular message passing. Driven by this mechanism, the influence of the mixed system is progressively applied to each molecule and atom, making message passing both efficient and meaningful. Comprehensive evaluations across diverse public datasets, spanning both pure components and mixtures, demonstrate that MesoNet achieves superior accuracy and enhanced chemical interpretability for molecular properties. This work establishes a powerful, interpretable approach for modeling compositional complexity, aiming to advance chemical simulation and design.*

## Introduction

Modern advancement in chemical and materials science demands efficient, accurate prediction of various physiochemical properties in complex systems, for which proper description of compositional interplay is crucial.[1-4] These intrasystem interactions matter for the deviation of a real system from an ideal one, and may exhibit in distinct material functions or chemical reactivities. While theories like quantum mechanics and classical thermodynamics provide fundamental insights, they often fail in mapping accurately the property evolution from molecular level to bulk systems.[5-6] Such deficiency limits the rational design of chemical systems toward advanced functions.

Data-driven machine learning (ML) techniques, especially deep learning, have emerged as a powerful paradigm to accelerate material discovery and chemical process design.[7-10] Through complex modeling of large datasets, ML models may correlate the structural features of molecule with their chemical behaviors,[11-18] among which the Graph Neural Networks (GNNs)[19,20] is outstanding. GNNs treat molecules as graphs, classifying atom and bond as node and edge, respectively, thus enabling efficient information collection and passing.[21-22] However, such information processing primarily focuses on intramolecular interactions and relies on non-linear transformations of atomic features and limited chemical descriptors (e.g., hybridization, valence).[23] Consequently, the GNN models lack interpretability and enough capacity to capture broad chemical context of complex systems.[17] To achieve a comprehensive understanding of molecular behavior, the model must be augmented with proper description of both intrinsic atomic properties and the compositional interplay in complex chemical environment. Indeed, more efficient message processing and passing policy is required to construct a more interpretable model.[24] However, the modeling of intrasystem interactions, especially their variants from atom to molecule, remains challenging.

Herein, we report the MesoNet, a framework for general, accurate prediction of physiochemical properties for complex systems based on multi-representation learning of chemical information. Thus, MesoNet augments the classic molecular graph with parameterized deep features of atom, which allow the integration of intramolecular and intermolecular interactions. This approach enables explicit, efficient information flow over different atoms, groups and molecules in complex systems. The core advancement of MesoNet lies in the definition of dynamic features that evolute upon encoding compositional interplay across contexts (e.g. from atomic to intermolecular interactions), allowing bidirectional information diffusion over molecules and their surrounding environment. This allows proper and efficient description of complex intrasystem interactions in multi-component systems with varying composition.

To grab the physiochemical properties and environmental context of each atom, MesoNet employs a hierarchical strategy capture the features and pass the message in a chemically meaningful manner. The information processing in mixture systems is thus explicit for different compositions. This results in significant improvements in the accuracy and interpretability of the model, offering a more effective approach for general property prediction of complex systems. By contrast, the previous models often struggle with mixtures. The framework of the model and the origins of its high performance will be discussed.

## The MesoNet Framework: Multi-Representation Learning for multi-molecule modeling.

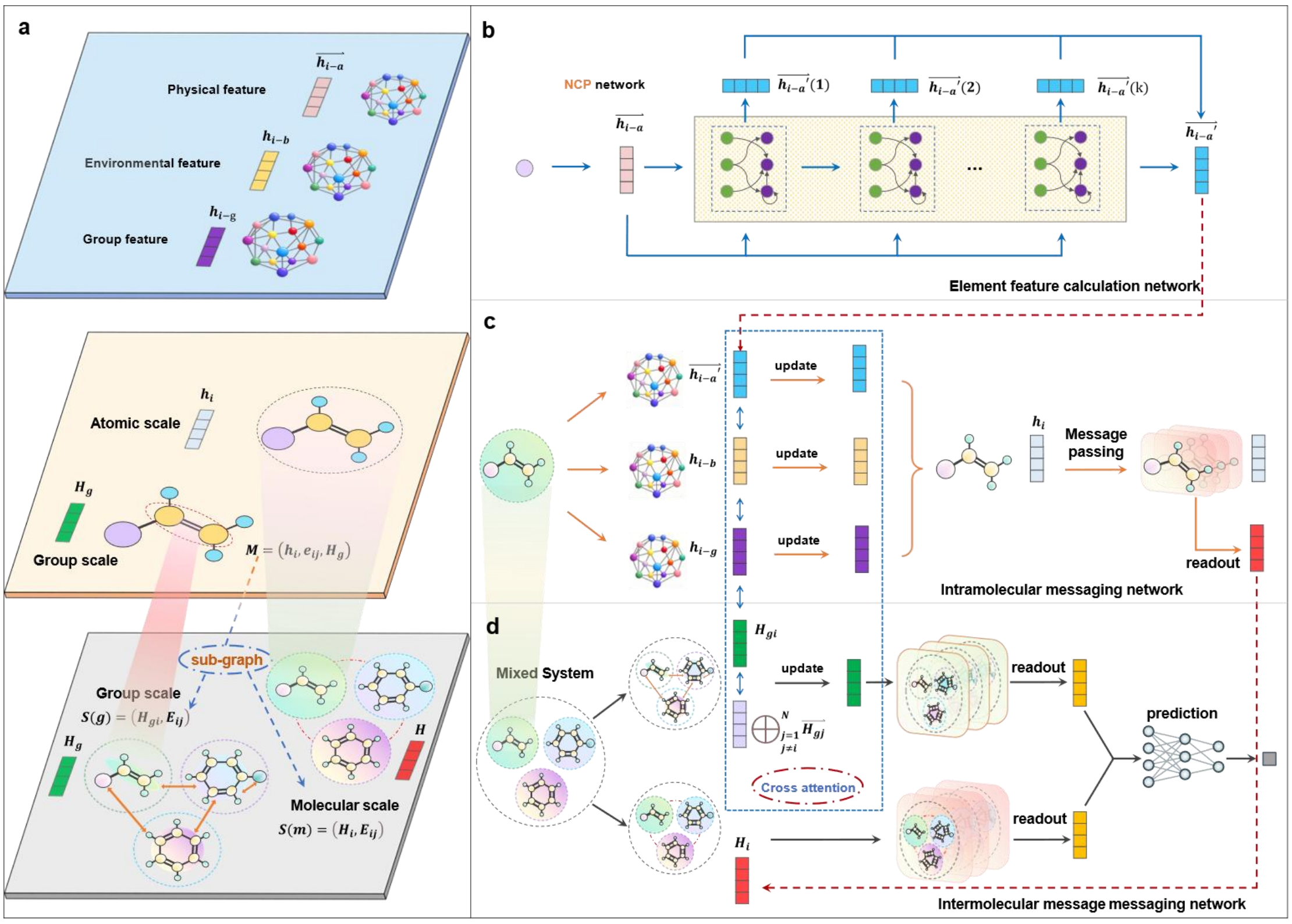

**Figure 1.** The MesoNet framework. Schematic representation for: (a) the hierarchical graph structure, including atomic physical feature graph, environmental feature graph, molecular graph, and mixture graph; (b) Parameterization of elemental features by employing Neural Circuit Policies (NCPs); (c) the intramolecular message passing within molecular graphs using a cross-attention based messaging network; (d) the intermolecular message passing between molecular graphs within the mixture.

To describe the true states of each atom, functional group and molecule in specific chemical environments, as well as the complex interplay therein, we define two graphs for the mixture system,

$S = (H_i, E_{ij}), Sg = (H_{gi}, E_{ij})$,

together with four molecular subgraphs:

$M(g) = (h_{i-g}, e_{ij})$, $M(b) = (h_{i-b}, e_{ij})$, $M(a) = (h_{i-a}, e_{ij})$ and $M = (h_i, e_{ij}, H_g)$ (Fig1.a).

Here, $H_i$ denotes the molecular feature of the mixture, while $H_{gi}$ encodes the group feature. At the atomic level, $h_{i-g}$ (group assignment), $h_{i-b}$ (environment) and $h_{i-a}$ (intrinsic feature) represent three complementary atomic descriptors; $h_{i-g}$ shares the same coding scheme as the global feature $H_g$ to facilitate message passing.

Next, we design a framework to share and transform feature at a molecular level, thereby enabling interactions among molecular graphs. To this end, three modules are employed:

**Elemental Feature Extraction Module：** This module generates intrinsic physical descriptors for each atom and match them with other atomic descriptors (Figure1a). These physically informed features serve as the initialization of the parameterization process, ensuring that atomic representations are not derived from random embeddings but grounded in physical and chemical knowledges.

**Message passing：** The intramolecular message-passing stage fuses diversified descriptors of an atom and propagates these information to its neighbours. The intermolecular stage performs message passing over groups and molecules to complement feature description with non-bonding interactions (Figure1b and 1c).

**Cross-Attention Module:** In this module the above extracted features interact with each other to simulate compositional interplay spanning from atom to molecules. Thus, each atom and molecule may announce their chemical environment even prior to message passing (Figure 1b and 1c).

Based on the above framework, the so abstracted atomic features are defined as Context-Aware Representation (CAR) parameters.

To drive the above framework efficiently, feature engineering is crucial. The Neural Circuit Policy (NCP)[25-26] is thus incorporated to extract deep atomic features through a neural architecture resembling Caenorhabditis elegans. To the best of our knowledge, such an architecture has not been employed in chemical modeling tasks before, which affords a novel application of natural rules in molecular representation learning. Unlike standard linear or attention-based transformations, NCP employs a sparsely connected recurrent process with gating mechanisms; this selectively amplify or suppress input signals, thereby offering a refined characterization of deep atomic features. In the MesoNet framework, the term *dynamic* refers to, not temporal dynamics, but rather the context-dependent adaptation of atomic descriptors during the course of message passing and cross-attention. This enables the extracted features to remain chemically meaningful while being adaptively refined in harmony with complex compositional environments. The model can thus parameterize abstract information, e.g. atomic electronegativity, first ionization energy, and covalent radius, and store it in the CAR parameters as defined above. The complex interactions are thoroughly considered using the following computation at a single neuron:

$$\overrightarrow{h_t} = \sigma\left(f\left(\left[\overrightarrow{h_{\iota-a}}, \overrightarrow{h_{t-1}}\right]\right)\right) \odot \sigma\left(g\left(\left[\overrightarrow{h_{\iota-a}}, \overrightarrow{h_{t-1}}\right]\right)\right) + \left[1 - \sigma\left(f\left(\left[\overrightarrow{h_{\iota-a}}, \overrightarrow{h_{t-1}}\right]\right)\right)\right] \odot \sigma\left(h\left(\left[\overrightarrow{h_{\iota-a}}, \overrightarrow{h_{t-1}}\right]\right)\right)$$

in which $\sigma$ denotes the activation function, while $f$, $g$, and $h$ are learnable transformation functions. This equation performs initial transformation of the atomic physical feature $\overrightarrow{h_{\iota-a}}$ and introduces a recurrent dependency on the previous time-step feature $\overrightarrow{h_{t-1}}$. The evolution of atomic features over time can thus be modeled, enabling efficient learning of dynamic interactions within the molecular structure.

Subsequently, the CAR parameters are refined as follows:

$$\overrightarrow{h_{\imath-a}}'(t) = NCP(\overrightarrow{h_{\imath-a}}, \overrightarrow{h_{\imath-a}}'(t-1))$$

during which CAR parameters are recursively derived through an NCP network. This is achieved using multiple Closed-form Continuous-time (CfC)[25] Model. Finally, the features extracted from different time spans are combined to comprehend the temporal evolution of atomic interactions:

$$\overrightarrow{h_{\imath-a}}' = concat(\overrightarrow{h_{\imath-a}}'(1), \overrightarrow{h_{\imath-a}}'(2) \dots \dots \overrightarrow{h_{\imath-a}}'(t))$$

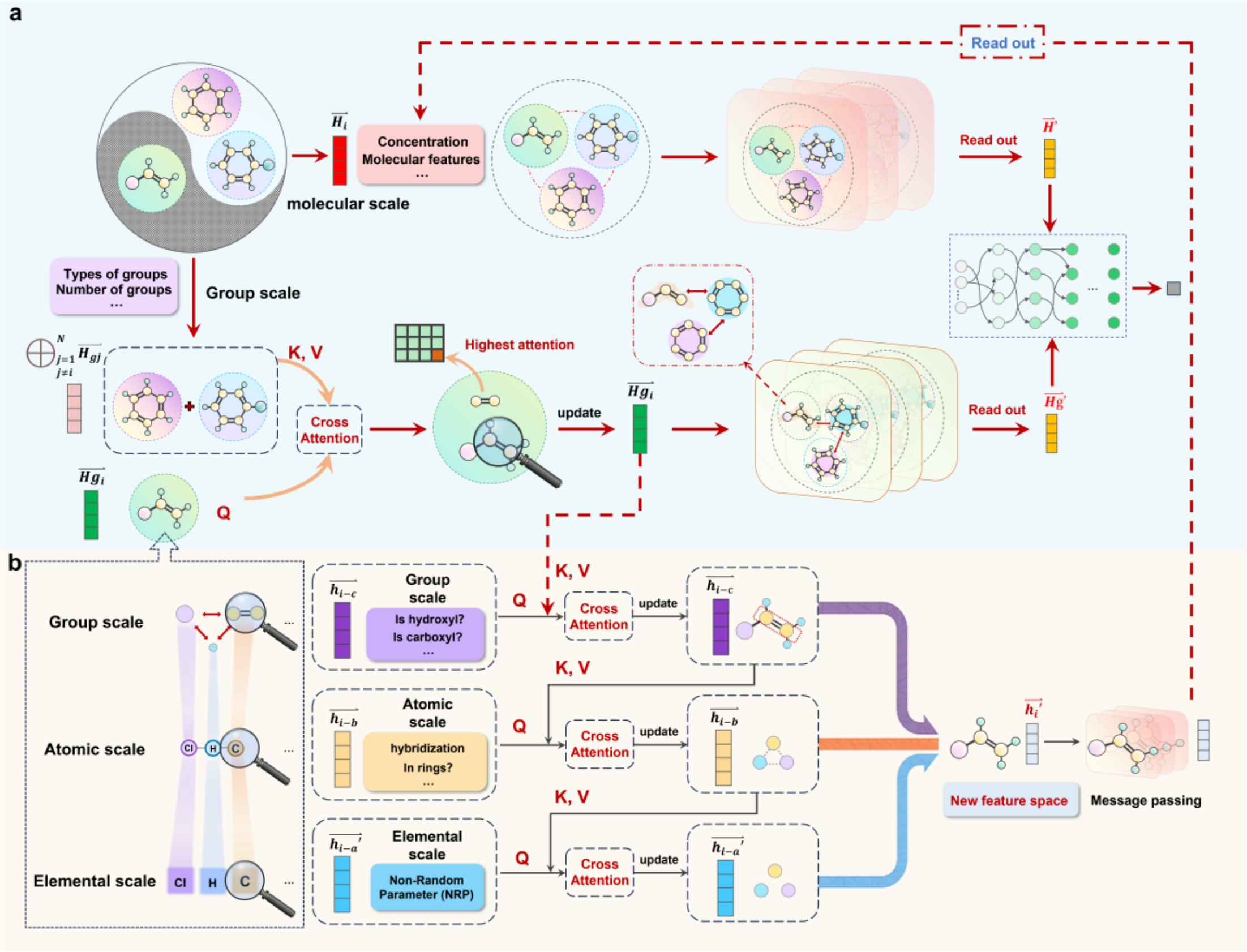

**Figure 2.** Flow chart of the cross-attention module. Schematic diagram of (a) intermolecular and (b) intramolecular message transmission and the associated cross-attention processes, respectively.

As shown in Figure 2a and 2b, in prior to message passing, the features are subjected to specific chemical environments via cross-attention, which is divided into

four modules:

**Mixture → molecule (group level)**:

$$Q = \overrightarrow{H_{gi}} * W_Q \ , \ K = \bigoplus_{\substack{j=1 \\ j\neq i}}^{N} concat(\overrightarrow{H_{gj}}, C_{gj}) * W_K, V = \bigoplus_{\substack{j=1 \\ j\neq i}}^{N} concat(\overrightarrow{H_{gj}}, C_{gj}) * W_V$$

$$Attention = softmax\left(\frac{QK^T}{\sqrt{d_k}}\right) * V$$

$$\overrightarrow{H_{gi}} = LayerNorm(Q + Attention)$$

Here, $W_Q$, $W_K$ and $W_V$ denote the query, key, and value matrices used in computing the cross-attention mechanism, respectively; $C_{gj}$ represents the concentration of neighboring molecules.

**Global group → atomic group assignment:**

$$\overrightarrow{h_{i-g}} = LayerNorm(Q + softmax\left(\frac{QK^T}{\sqrt{d_k}}\right) * V)$$

where $Q = \overrightarrow{h_{i-g}} * W_Q$, $K = \overrightarrow{H_{gi}} * W_K$, $V = \overrightarrow{H_{gi}} * W_V$.

**Group assignment → atomic environment:**

$$\overrightarrow{h_{i-b}} = LayerNorm(Q + softmax\left(\frac{QK^T}{\sqrt{d_k}}\right) * V)$$

where $Q = \overrightarrow{h_{i-b}} * W_Q$, $K = \overrightarrow{h_{i-g}} * W_K$, $V = \overrightarrow{h_{i-g}} * W_V$.

**Environment → intrinsic physical features of atoms:**

$$\overrightarrow{h_{i-a}} = LayerNorm(Q + softmax\left(\frac{QK^T}{\sqrt{d_k}}\right) * V)$$

where $Q = \overrightarrow{h_{i-a}} * W_Q$ , $K = \overrightarrow{h_{i-b}} * W_K$ , $V = \overrightarrow{h_{i-b}} * W_V$ . The updated atomic representation thus becomes:

$$\overrightarrow{h_i} = concat(\overrightarrow{h_{i-a}}, \overrightarrow{h_{i-b}}, \overrightarrow{h_{i-g}})$$

Further, intramolecular message passing proceeds via:

$$\overrightarrow{h_i}^{k} = \Theta \overrightarrow{h_i}^{k-1} + \frac{1}{N} \sum_{j \in N(i)} \overrightarrow{h_j}^{k-1} * (\sigma(W * \overrightarrow{e_{ij}}))$$

which allows continuous update of atomic features upon integrating the information of neighboring atoms. The graph features are aggregated using a readout operation:[27]

$$\overrightarrow{H_i} = readout(\sum_{N(i)} \overrightarrow{h_i})$$

which aggregates the node features to constitute global features.

For efficient intermolecular message passing, the delivery of chemical information has to be properly delt with. Such information includes hydrophilicity, potential π-π stacking, hydrogen bonding capabilities, etc. Here, we employ two distinct message-passing mechanisms to capture both global and local contributions of each component (Figure 2):

**Molecular Level:** The contribution of different molecules under various concentrations is considered, and a few edge features are incorporated, e.g. hydrogen bond donors, acceptors, and topological polar surface area, as well as the influence of van der Waals forces and hydrogen bonding ($\overrightarrow{E_{ij}}$). The corresponding equation is as follows:

$$\overrightarrow{H_i}^{k} = \Theta \overrightarrow{H_i}^{k-1} + \frac{1}{N} \sum_{j \in N(i)} \overrightarrow{H_j}^{k-1} * (\sigma(W * \overrightarrow{E_{ij}}))$$

**Group Level:** After the cross-attention processes, the global eigenvectors of groups are used again. This quantifies both intra- and intermolecular group interactions:

$$\overrightarrow{Hg_i}^{k} = \Theta \overrightarrow{Hg_i}^{k-1} + \frac{1}{N} \sum_{j \in N(i)} \overrightarrow{Hg_j}^{k-1} * (\sigma(W * \overrightarrow{E_{ij}}))$$

Subsequently, the differential features upon subjecting the molecule in a mixture are considered, as reflected by two correction vectors:

$$\vec{H}' = readout(\sum_{N(i)} concat(\overrightarrow{H_i}, \overrightarrow{H_i}^{k}))$$

$$\overrightarrow{Hg}' = readout(\sum_{N(i)} concat(\overrightarrow{Hg_i}, \overrightarrow{Hg_i}^{k}))$$

During the above message coding and passing processes, the utilization of CAR parameters allows for quantification of abstract essence and principles of substance, affording explicit and efficient propagation of physiochemical information across the entire system.

Finally, a Multi-Layer Perceptron ($MLP$) model is employed to accomplish the final prediction, combining chemical and physical features to produce a robust estimation of physiochemical properties:

$$y = MLP(concat(\overrightarrow{H_i}, \overrightarrow{H'}, \overrightarrow{Hg'}))$$

Such a non-parameterized network, MesoNet, is thus constructed.

## Results and Discussion

To evaluate thoroughly the performance of MesoNet, a set of diverse datasets were employed, including water solubility, lipophilicity, ionization energy (IE), critical micelle concentration (CMC) of surfactants, spectral properties (e.g. specifically absorption wavelength, emission wavelength, and photoluminescence quantum yield (PLQY) measured across different solutions), and the activity coefficients of binary and ternary mixtures. The datasets for fixed-composition mixtures were categorized as pure components.

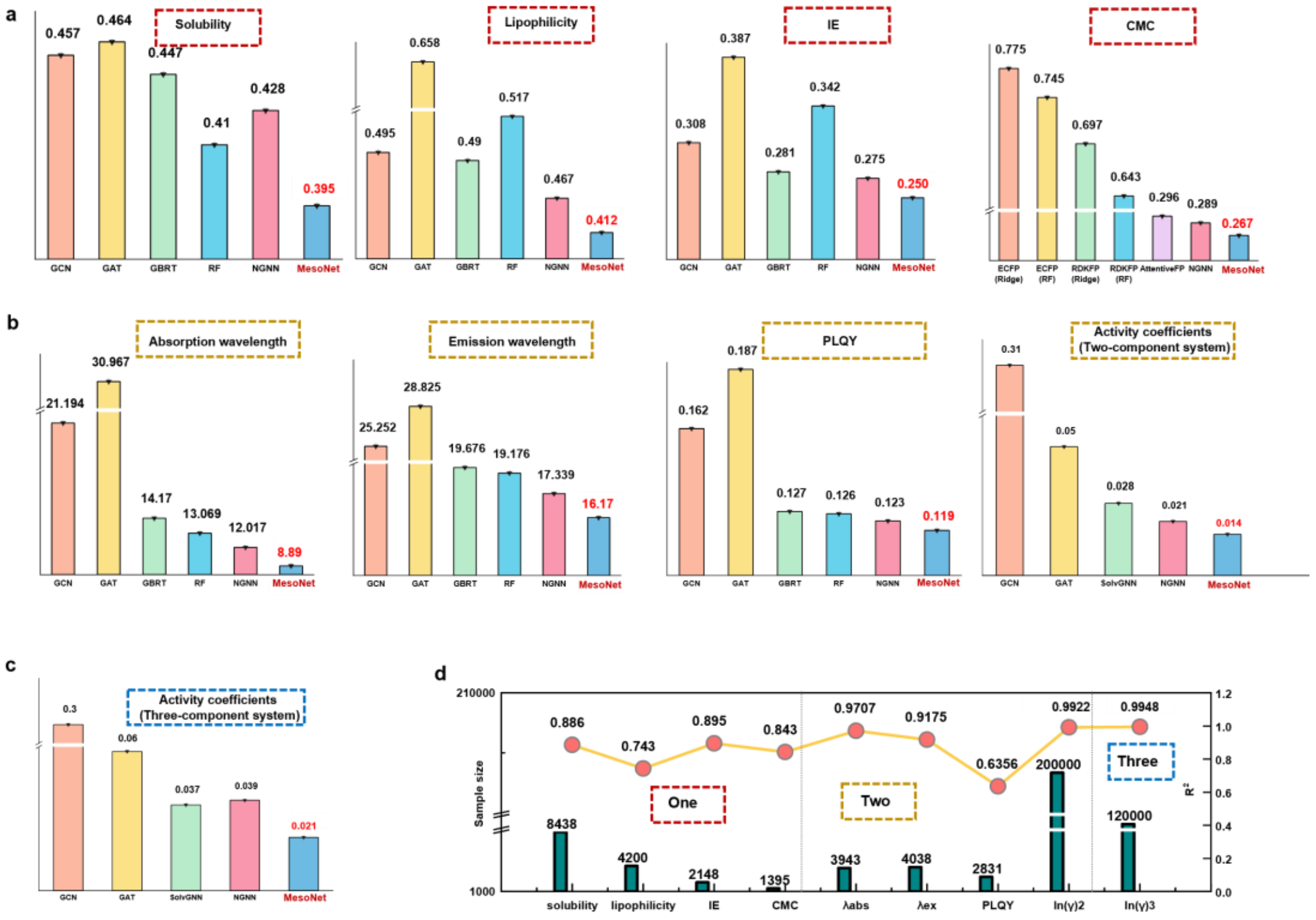

**Figure 3.** The performance of different models on the prediction of various properties for: (a) pure components, (b) binary systems, and (c) ternary mixtures; (d) the sample size of different datasets vs. the $R^2$ afforded by MesoNet.

**Table 1**. Prediction results of activity coefficients using different models.

| | Activity coefficient (two-component without Inf [b]) | | | Activity coefficient (two-component with Inf) | | | Activity coefficient （three-component） | | |
|---|---|---|---|---|---|---|---|---|---|
| | MAE | RMSE | $R^2$ | MAE | RMSE | $R^2$ | MAE | RMSE | $R^2$ |
| GDI -GNN[32] | - [a] | - | - | 0.028 | 0.081 | 0.990 | - | - | - |
| GDI-GNN*x*MLP[32] | - | - | - | 0.025 | 0.083 | 0.990 | - | - | - |
| GDI-MCM[32] | - | - | - | 0.030 | 0.088 | 0.989 | - | - | - |
| GE-GNN[31] | - | - | - | 0.020 | 0.068 | 0.993 | - | - | - |
| solvGNN[28] | 0.028 | 0.077 | 0.980 | 0.030 | 0.088 | 0.989 | 0.037 | 0.072 | 0.990 |
| NGNN[24] | 0.020 | 0.064 | 0.989 | 0.023 | 0.071 | 0.992 | 0.039 | 0.084 | 0.983 |
| **MesoNet** | **0.014** | **0.050** | **0.992** | **0.015** | **0.057** | **0.995** | **0.020** | **0.047** | **0.995** |

[a]: '-' indicates that the data is lacking; [b]:'Inf' indicates the infinite dilution activity coefficient data. The above results are under a five-fold cross validation .

In general, MesoNet outperformed previously reported models[24,28-32] over diverse datasets. As illustrated in Figure 3, MesoNet kept robust performance in the prediction of various properties across pure components and mixtures. Indeed, during the training of intermolecular message passing, while the intrinsic features of each component maintain, the quantified interactions between molecules are optimized. This prevails for all tested systems, especially in the prediction of spectral properties (Figure 3b) and activity coefficients (Figure 3c), where MesoNet achieved remarkably lower MAE and tighter error bars. Furthermore, Figure 3d reveals a positive correlation between the number of components and the accuracy of MesoNet in predicting activity coefficients. MesoNet's superiority in comprehending compositional interplay is thus validated.

Further, as shown in Table 1, MesoNet demonstrated exceptional accuracy in predicting activity coefficients. Moreover, it is better at describing the *p*-*V*-*T* relation of the mixture system, as reflected by the high $R^2$ value even without the infinite dilution activity coefficient data. Note that the design of MesoNet does not apply thermodynamic constraints[31-32] nor does it specifically handle solvent effects,[24] thus indicating high generalizability of the model in learning and exerting physiochemical principles.

To further validate the ability of MesoNet in handling intermolecular interactions, the binary and ternary activity coefficient datasets were merged for test (as detailed in the Method section). As shown in Figures 4a to 4c, on the mixed dataset, MesoNet converged faster with higher accuracy and better generalization. Apparently, higher complexity of the system even facilitates the model in learning the compositional interplay.

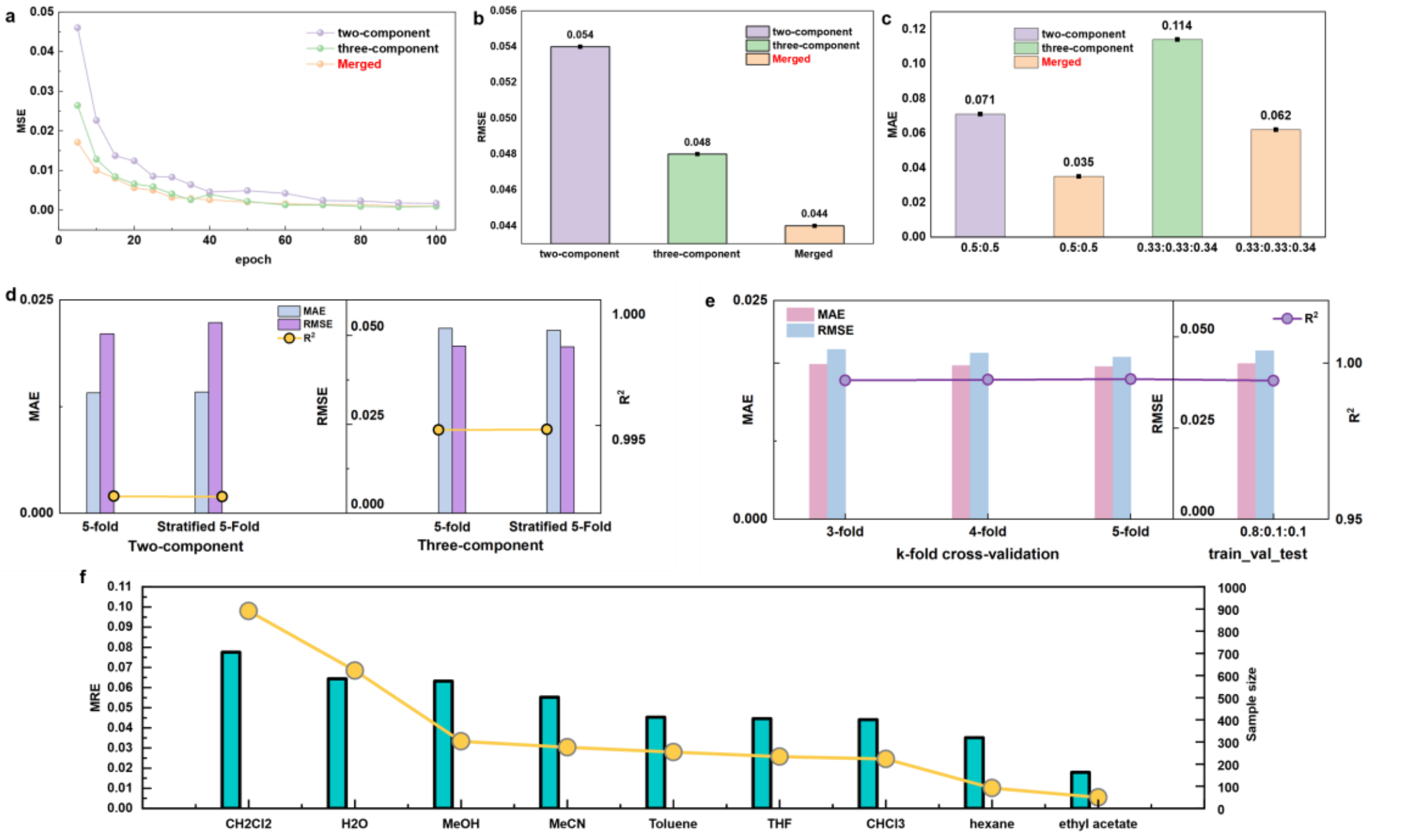

**Figure 4.** The performance of MesoNet in handling intermolecular interactions. (a) The training curves, (b) error bars, (c) the extrapolation (to different concentrations) results, and (d) prediction results with different sampling strategies with different activity coefficient datasets; (e) prediction results with different dataset partitioning method; (f) The prediction results for optical properties with unprecedented solvents (Hexane, Ethyl Acetate, THF, $CHCl_3$).

Next, as shown in Figure 4d, both five-fold random cross-validation and five-fold stratified cross-validation yielded consistent results (MAE, RMSE, $R^2$), demonstrating the robustness of the MesoNet framework in learning intermolecular interactions. Further evidence was provided by the extrapolation experiments: as shown in Figure 4f, upon increasing the size of the training set, the error progressively decreased, confirming that the model can generalize solvent effects for unseen solvents.

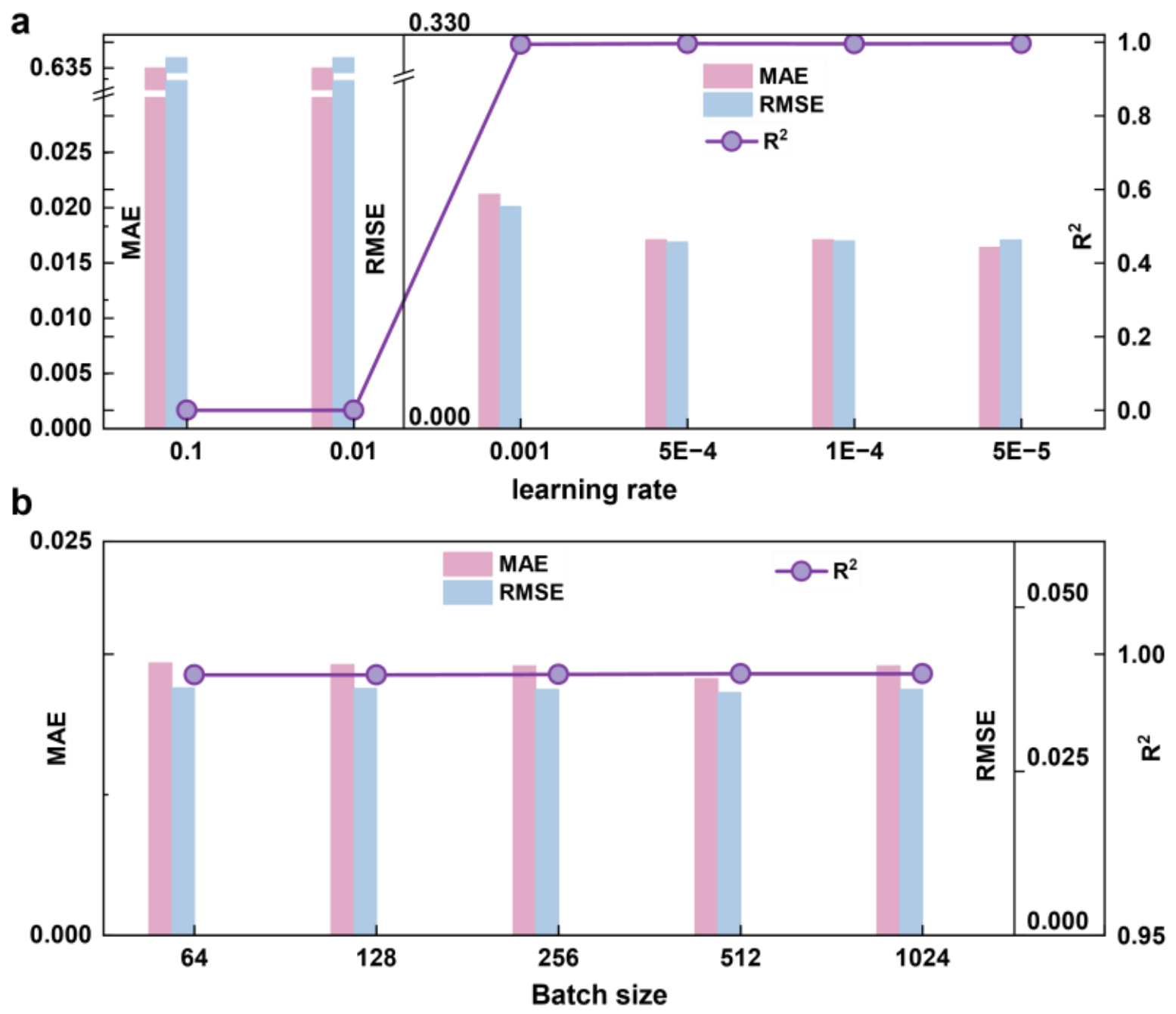

**Figure 5.** The influence of hyperparameters on the prediction results of mixed multi-component activity coefficients (a) learning rate; (b) batch size.

Notably, we also investigated the effect of selected hyperparameters on model behavior. As shown in Figure 5, excessively high learning rates cause convergence failure. By contrast, with even a rather low learning rate (e.g. $5e^{-5}$), the performance of the model remains stable, indicative for inherent robustness. In panel (b), the choice of batch size exerts little influence on the final prediction accuracy, suggesting that MesoNet is capable of fast learning.

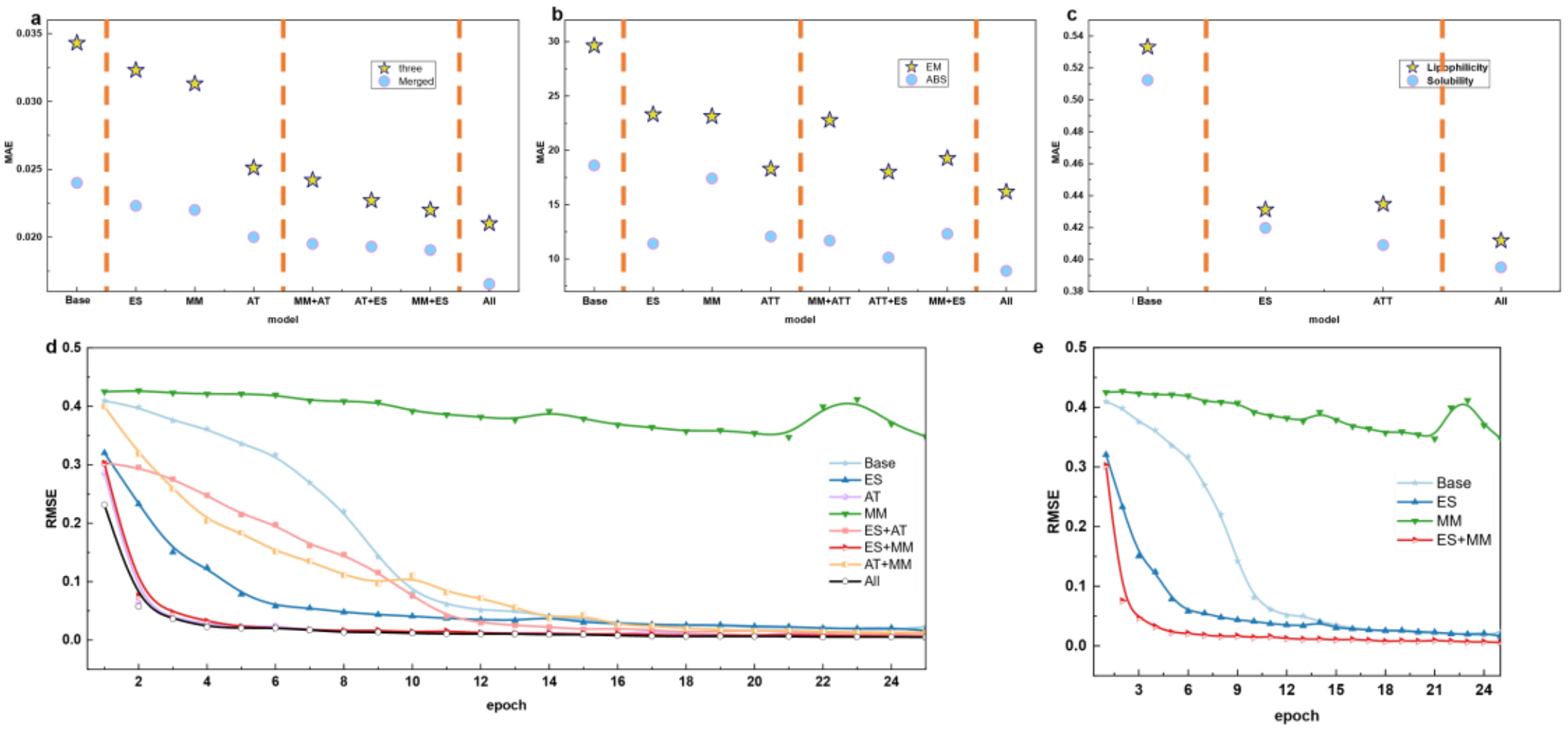

**Figure 6.** The results ablation study. The performance of the model upon adding different modules for the prediction of: (a) activity coefficients, (b) absorption (ABS) and emission wavelengths (EM), (c) solubility and lipophilicity; (d) the training curves of the model upon adding different modules for the prediction of activity coefficients; (e) the impact of feature extraction modules on training speed. Note: Base represent the baseline model without any added modules; MM represents the intermolecular message passing module; ATT represents the cross-attention module; ES represents the element feature extraction network.

**Ablation Studies.** A systematic ablation study was conducted to investigate how different modules contribute to the performance of MesoNet. A message-passing neural network (MPNN) was used as the basic module. As shown in Figures 5a – 5c, while affording unsatisfactory results in the prediction of activity coefficients, lipophilicity and solubility, the performance of MPNN was improved gradually upon adding different modules. However, for the prediction of absorption and emission wavelengths, different modules concert with each other, and the performance of the model can be improved significantly only by incorporating all modules.

**Training Speed Analysis**. The impact of different modules on the training speed has been examined. As shown in Figure 5d, adding only the intermolecular message-passing module (MM) resulted in slower convergence compared to the baseline model.

By contrast, upon incorporating the feature extraction module (ES), the convergence was accelerated dramatically (Figure 5e). Moreover, the ES may well concert with MM thus affording even faster convergence, indicating that the CAR parameters serve as adaptive parameters favoring both the feature capture and the message passing.

Based on the above results, it is also interesting to figure out how the CAR values are assigned, which is crucial for the excellent performance of MesoNet. As shown in Figures 7a – 7b, in the prediction of activity coefficient for ternary systems, the model classified elements into hydrogen and non-hydrogen groups, while the non-hydrogen elements are still distinguishable within a narrow value scale. It turned out that the network is able to capture the importance of each element and grab their subtle differences.

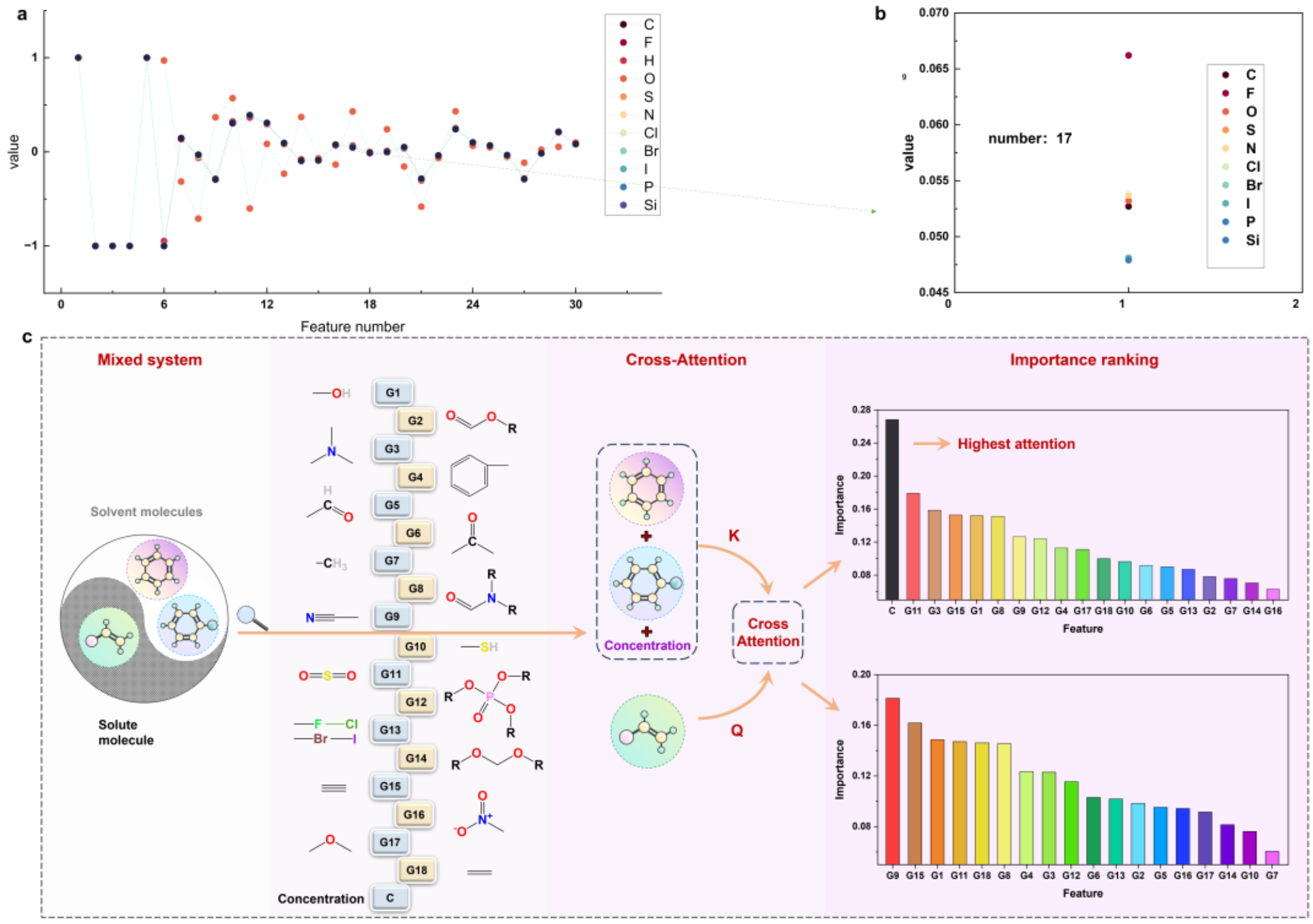

**Figure 7.** Analysis of the CAR features. (a)The numerical values of 30 features of each element extracted with NCPs; (b) the values of the 17th feature for different elements; (c) the attention weights assigned by the intermolecular cross-attention module to the functional groups of solute and solvent molecules.

Notably, we also observed that when employing the cross-attention module alone, although the prediction accuracy is lower compared to the combined architecture (6a), the training process converges significantly faster compared to models with other individual modules added (6d). To further investigate this phenomenon, we analyzed the behavior of cross-attention during intermolecular message passing. As illustrated in Figure 7c, the attention weights assigned to environmental molecules reveal that the concentration term receives disproportionately high weights, far exceeding those of group-level descriptors. This indicates that the cross-attention module can directly identify and prioritize critical factors for mixture systems—particularly concentration—at an early stage. By injecting this information into the representations prior to message passing, the network accelerates convergence, as it no longer needs to relearn from scratch the joint influence of molecular structure and concentration during intermolecular message propagation.

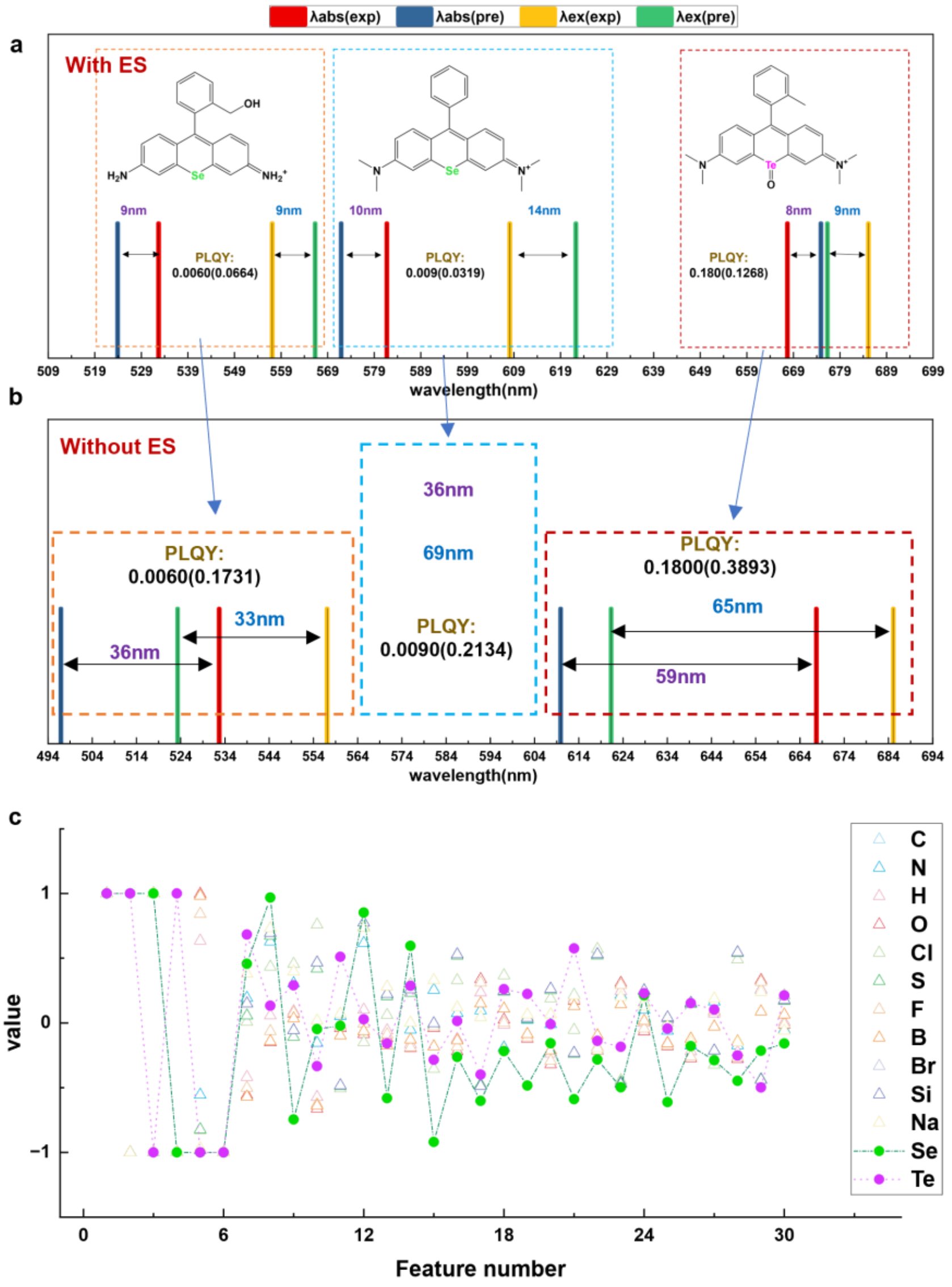

**Figure 8.** Extrapolation of the prediction to unprecedented elements. (a) The prediction of optical properties for molecules containing Te and Se that have not been included in training by using (a) MesoNet and (b) the one ablating the ES module; (c) Visualization of feature distributions for extrapolated (unseen) elements (Te, Se) and trained elements, as extracted by the atomic feature extraction network.

For physiochemical property prediction, the generalizability of the model is also reflected in whether or not the extrapolated property for untrained elements is reliable. Here, as shown in Figure 6a, for the optical properties of molecules containing tellurium (Te) and selenium (Se) (unprecedented in the training set), high prediction accuracy was achieved with MesoNet. By contrast, when extruding the CAR parameters by ablating the ES module, the accuracy dropped sharply (Figure 6b). Further, the extrapolated CAR values for Te and Se exhibited distinguishable patterns rather than collapsing into identical values (Figure 6c), such differentiation is essential for generalization, as identical values across distinct elements would indicate a failure to learn new chemical information, this indicating powerful ability of MesoNet in comprehending the nature of element.

## Conclusion

In summary, the MesoNet framework represents a significant advancement in molecular property prediction, overcoming the limitations of existing approaches by introducing a new paradigm for modeling complex chemical systems. The motivation for this work was to address the challenge for conventional models like GNNs, which rely on static, predefined features but cannot effectively capture sufficient compositional interplay in complex systems. To this end, we introduced the core concept, i.e. context-aware representation – dynamically adaptive descriptors which reflect both the essential feature of an atom and its environmental context. This approach constructs a richer and more informative feature space.

To implement this concept, MesoNet employs a unique hierarchical architecture. That is, a multi-level message-passing mechanism concerts with a cross-attention module to facilitate efficient information flow across atomic, functional group, and molecular levels. To the best of our knowledge, such architecture achieves, for the first time, a unified modelling of intra- and intermolecular interactions. Rigorous validation on diverse datasets, including solubility, activity coefficients, and spectral properties, has confirmed the exceptional accuracy and generalizability of MesoNet.

The construction of CAR parameters offers a new perspective on the representation and modeling of chemical information. We anticipate that it can bridge the gap between traditional physics-based models and "black-box" machine learning approaches, enhancing both the interpretability and the active-learning capability of the model. At the current stage, this has been validated primarily on datasets composed of organic molecules, including pure components and mixtures. While organometallic compounds remain outside the present scope, there is no limitation in principle when associated datasets are available. On the other hand, for extended property prediction with diverse chemical context, this principle-driven framework can be conveniently transferred and combined with other functional modules.

## Data And Method

### Graph Representation of Mixed Systems

In MesoNet, each molecule in two mixed system( $S = (H_i, E_{ij})\ and\ Sg = (H_{gi}, E_{ij})$) is represented as an individual subgraph, with atoms from different molecules distinguished by a masking mechanism. The molecular-level features ($H_i$) are extracted via a readout function applied after the message-passing phase and then concatenated with the molecule's global features, The group level features ($H_{gi}$) are added in advance to the global variables of each subgraph molecule by extracting molecular substructures， and used for the subsequent calculation of cross attention。

Intermolecular features ($E_{ij}$) incorporated into the model include:

(1) hydrogen bond donors

(2) hydrogen bond acceptors

(3) topological polar surface area

(4) water–oil partition coefficient (LogP).

Each molecular graph is further decomposed into three graphs to capture complementary aspects of atomic features. Each subgraph will have a global feature: the group-level feature **(**$H_g$**)**. This feature is computed using SMARTS during the creation of the molecular graph and includes the types and quantities of common functional groups. The global vector is then employed in cross-attention with both the remaining molecules in the mixture and the group encodings of individual atoms (h_{i-g}), which are represented in a consistent manner.

The physical feature graph ($M(a) = (h_{i-a}, e_{ij})$) encodes intrinsic properties of each atom ($h_{i-a}$), such as:

(1) atomic number

(2) electronegativity

(3) covalent radius

(4) atomic mass

(5) first ionization energy

(6) electron affinity.

In parallel, the environmental graph ($M(b) = (h_{i-b}, e_{ij})$) captures the contextual and configurational attributes of each atom, including its hybridization, valence state, and directional properties. The atomic features ($h_{i-b}$) considered are summarized as follows:

(1) Symbol: {B, Br, C, Cl, F, Ge, H, I, N, Na, O, P, S, Se, Si, Te}

(2) Number of Valence Electrons: {0, 1, 2, 3, 4, 5, 6}

(3) Number of Hydrogens: {0, 1, 2, 3, 4}

(4) Formal Charge: {-2, -1, 0, 1, 2}

(5) Hybridization: {s, sp, $sp^2$, $sp^3$}

(6) Ring Membership (isInRing): {True, False}

(7) Aromaticity (isAromatic): {True, False}

Both graphs share common edge features ($e_{ij}$), which encode the bonding interactions between atoms:

(1) Bond Type: {single, double, triple, aromatic}

(2) Conjugation: {True, False}

(3) Aromaticity: {True, False}

(4) Ring Bond: {True, False}

Each subgraph will have a global feature: the group-level feature $(H_{gi})$. This feature is computed using SMARTS during the creation of the molecular graph and includes the types and quantities of common functional groups.

This comprehensive graph representation allows the model to effectively capture both intra- and intermolecular interactions, thereby providing a more accurate description of the complex chemical environment in mixed systems.

**Data and Training**

We used data sourced from the literature,[24,28-30] to better compare the performance of our model, we compared the results with those from the best-performing models and baseline models reported in the literature. Except for activity coefficient prediction, which was evaluated using five-fold cross-validation, all other datasets were split into training, validation, and test sets with a ratio of 0.8:0.1:0.1. In addition, we compared in the main text the results obtained from five-fold cross-validation and from the 0.8:0.1:0.1 split for the activity coefficient dataset. The results show that even when selecting the best-performing test set from the five-fold cross-validation, the accuracy is comparable to that achieved with the 0.8:0.1:0.1 split. This consistency indicates that the model exhibits strong generalization ability.

**Solubility Prediction**: The results were compared with the baseline models reported in the literature, including GCN, GAT, RF, GBRT and NGNN, [24] as well as the best-performing model (RF). [30]

**Critical Micelle Concentration (CMC) Prediction**: We compared our model's performance against baseline models (ECFP-RF, ECFP-Ridge, RDKFP-RF, RDKFP-Ridge) and the best-performing model (AttentiveFP). [29]

**Lipophilicity and Ionization Energy (IE) Prediction**: The results were compared

with the baseline models (GCN, GAT, RF, and GBRT) and the best-performing model (NGNN) as reported in the literature.[24]

**Two-component Spectral Properties Prediction:** For the absorption wavelength, emission wavelength, and Photoluminescence Quantum Yield (PLQY) predictions, the data were sourced from website (http://www.chemfluor.top). Results were compared with baseline models (GCN, GAT, RF, GBRT) and the best-performing model (NGNN) as reported in the literature. [24]

**Two-component and Three-component Activity Coefficients Activity Coefficients Prediction**: For the predictions of activity coefficients without infinite dilution, the methodology followed in the literature was consistent, [28] using five-fold cross-validation. We compared the performance of our model with baseline models (GCN, GAT) and best-performing model (solvGNN). [28] For predictions of activity coefficients with infinite dilution, we compared our model's performance with the latest models reported in the literature, including GDI-GNN34[32], GDI-GNNxMLP34[32], GDI-MCM34[32], GE-GNN33[31], solvGNN[28], and NGNN[24].

**Merged Activity Coefficients：** The dataset integrates binary and ternary activity coefficient data by augmenting the binary systems with a nitrogen molecule at zero concentration, thereby converting them into ternary representations.To evaluate the model's prediction performance, we used the following indicators: mean relative error (MRE), mean absolute error (MAE), root mean square error (RMSE), and the coefficient of determination ($R^2$). Further information about the data and framework is also available at https://github.com/Fan1ing/MesoNet.

## Acknowledgement

Generous financial support by the National Natural Science Foundation of China (U24A20527) and the Key Research and Development Program of Zhejiang Province (2023C01102, 2023C01208).